# On the distribution of time-to-proof of mathematical conjectures

Ryohei Hisano and Didier Sornette

ETH Zurich, D-MTEC, Kreuzplatz 5
CH-8032 Zurich, Switzerland
Emails: em072010@yahoo.co.jp and dsornette@ethz.ch

***Abstract***: What is the productivity of Science? Can we measure an evolution of the production of mathematicians over history? Can we predict the waiting time till the proof of a challenging conjecture such as the P-versus-NP problem? Motivated by these questions, we revisit a suggestion published recently and debated in the "New Scientist" that the historical distribution of time-to-proof's, i.e., of waiting times between formulation of a mathematical conjecture and its proof, can be quantified and gives meaningful insights in the future development of still open conjectures. We find however evidence that the mathematical process of creation is too much non-stationary, with too little data and constraints, to allow for a meaningful conclusion. In particular, the approximate unsteady exponential growth of human population, and arguably that of mathematicians, essentially hides the true distribution. Another issue is the incompleteness of the dataset available. In conclusion we cannot really reject the simplest model of an exponential rate of conjecture proof with a rate of 0.01/year for the dataset that we have studied, translating into an average waiting time to proof of 100 years. We hope that the presented methodology, combining the mathematics of recurrent processes, linking proved and still open conjectures, with different empirical constraints, will be useful for other similar investigations probing the productivity associated with mankind growth and creativity.

# 1. Introduction

In the December 25, 2010 issue of New Scientist (www.newscientist.com), a series of articles presented measures of scientific progress to predict the timing of new discoveries and to make some forecasts for the future of science and technology. In particular, Arbesman and Courtland used the waiting times that have been required to solve 18 mathematical problems to estimate that the probability for the "P versus NP problem" to be solved by the year 2024 is roughly 50% [1]. The methodology for this estimation consists essentially on the approximate construction of the cumulative distribution of waiting times from formulation to proof (referred to as "times-to-proof" thereafter), based on the above mentioned set of 18 solved mathematical problems.

Issues involving the methodology of their analysis created a heated debate in the comment space of the New Scientists' website [2]. For instance, one comment posted on February 3, 2010 10:27:11 GMT criticized the authors' methodology, stressing that "their method of estimation looked only at problems that actually were solved," which may introduce a selection bias. Another comment, which followed, replied to this criticism stating that "The method is certainly not flawed, as you don't have the contextual basis with which to make assignment of truth or falsity in this case..." and so on. A more formal attack to the column was published in the New Scientist issue of February 2, 2011 [3]. In a nutshell, its argument is that using a probability distribution amounts to assuming that the underlying generating process is stationary, but stationarity may not hold over the decades and centuries corresponding to the investigated data, given that the population of mathematicians has grown significantly and their theorem-proving technology has arguably improved due, e.g., to cumulative knowledge, computers, and collective work mediated by Internet and social network tools.

We revisit this question and analyze a larger database of 144 conjectures including both closed and open conjectures (instead of the limited 18 problems in [1]). We dissect the major problems that plague any attempt to infer the distribution of times-to-proof for mathematical conjectures, given the available data and the intrinsic non-stationarity of the system. We show that a large part of the database can be reasonably accounted for by an underlying time-to-proof distribution that is close to an exponential distribution with rate 0.01/year (translating into an average waiting time to proof of 100 years).

The next section 2 presents the dataset. Section 3 reviews and adapts the theory of recurrence processes to the distribution of time-to-proof. Section 4 presents the empirical distribution of time-to-proof obtained from our dataset. Section 5 formulates the consequences of the non-stationarity of the births of conjectures. Section 6 combines this non-stationarity with different models for the intrinsic distribution of time-to-proof to fit the empirical distributions. Section 7 concludes by stressing the caveats of the proposed analysis.

2. Dataset

In order to address the crucial question of the stationarity of the distribution of times-to-proof, the dataset of 18 problems used by Arbesman and Courtland [1] is insufficient. We thus turn to the dataset that the mathematical community has collectively contributed in constructing the page "list of conjectures" in Wikipedia, the free web encyclopedia, which lists about 160 proved and unsolved conjectures [4]. While it is difficult if not impossible to establish that this list is representative and does not represent a biased sample, other tests have shown that the accuracy of Wikipedia's articles compares well with that of the standard, Encyclopedia Britannica [5]. In absence of other data sources, our stance is that it is better to try to work with what is available than to do nothing. At least, we may learn the limits to overcome.

3 Distribution of time-to-proof: definition and theory

For each conjecture $i$ present in the "list of conjectures" in Wikipedia, we searched for the exact year $t_1^i$ when it was stated and the exact year $t_2^i$ when it was resolved (or whether it still remains open), always striving to obtain the first original source or reference. For 16 conjectures, we were unable to determine the exact values of $t_1^i$ and/or $t_2^i$, thus reducing our usable dataset to 144 conjectures, of which 60 have been solved until present (Jan 2012) and 84 are still open problems.

We determine the time-to-proof $\tau_I^c$ for each of the 60 conjectures that have been solved (or proven wrong) according to the formula

$$\tau_I^c := t_2^i - t_1^i \tag{1}$$

For the 84 open conjectures, the relevant variables are the so-called "backward recurrence times" defined by

$$\tau_I^b := t - t_1^i \tag{2}$$

where $t$ is the present (2012). Our strategy is to study the complementary cumulative distribution functions (ccdf) (also called "survivor functions") $S_c(\tau)$ and $S_b(\tau)$ corresponding to the times-to-proof $\tau_t^c$ and $\tau_t^b$ respectively.

The mathematical theory of interval distributions for stationary point processes provides an exact correspondence between the survival function of closed conjecture and open conjecture that we can use when interpreting the empirical distributions. Defining $N(\tau, \tau+t]$ as the number of events in the time interval $(\tau, \tau+t]$ (which excludes the left side and includes the right side of the time interval), the backward recurrence time of an event generated by a point process, is defined formally as

$$\tau_t := \inf\{u > 0 : N(t-u, t] > 0\} \qquad (3)$$

where $t = 2012$ (i.e. present year) [10]. In words, it is the time interval from the latest event (the formulation of a conjecture) to present (at which time the conjecture is still open), such that there is one event in this interval. For stationary process, we have the identity [11]

$$Pr\{N(0, \tau) \geq 1, N(\tau, \tau+t)=0\} = Pr\{N(\tau, \tau+t)=0\} - Pr\{N(0, \tau+t)=0\} \qquad (4)$$

In words, the probability $Pr\{N(\tau, \tau+t)=0\}$ that there are no events in $(\tau, \tau+t)$ is equal to the probability $Pr\{N(0, \tau+t)=0\}$ that there are no event in $(0, \tau+t)$ plus the probability $Pr\{N(0, \tau) \geq 1, N(\tau, \tau+t)=0\}$ that there are no events in $(\tau, \tau+t)$ and at the same time there is at least one event in $(0, \tau)$. In other words, the fact that the interval $(\tau, \tau+t)$ has no event can be associated with the occurrence of either no event or of some events earlier in $(0, \tau)$. Dividing both sides by $\tau$, taking the limit $\tau \Rightarrow 0$ of expression (4) and using the definitions

$$S_c(t) = \lim_{\tau \Rightarrow 0} Pr\{N(\tau, \tau+t)=0 \mid N(0, \tau) \geq 1\} \qquad (5)$$

and

$$S_b(t) = Pr\{N(0, t)=0\} \qquad (6)$$

for the complementary cumulative distribution functions $S_c(\tau)$ and $S_b(\tau)$ corresponding to the times-to-proof $\tau_t^c$ and $\tau_t^b$ respectively defined by (1) and (2), identity (5) translated into the Palm-Khinchin relation [10-12]

$$S_b(t) = -\frac{1}{\lambda}\frac{d}{dt}S_c(t) \tag{7}$$

where $\lambda$ is the inverse of the average time-to-proof (i.e. $\lambda = lim_{\tau=>0}\ Pr\{N(0, \tau) = 1\}/\tau$). The conditioning in (5) ensures that the counting of the time to the next event is indeed starting from the previous one (the condition $N(0, \tau) \geq 1$).

## 4 Empirical distributions of time-to-proof

It is important to note that the birth flow of mathematical conjectures is not uniformly distributed with time. For the dataset of 60 solved conjectures, fig. 1 shows a scatter plot in which each symbol corresponds to a given conjecture, with the abscissa giving the year when the conjecture was stated and the ordinate giving the year when the conjecture was solved. The two time axes cover both the period from 1600 CE to present. We can see that the plot is significantly crowded at the upper right part of the figure. This implies that the birth flow of conjectures is increasing with time.

Fig. 2 quantifies this visual impression by showing the cumulative number of stated problems $N_{closed}(t)$ (that have found a solution) and $N_{stated}(t)$ (that are both closed and still open) and the cumulative number $N_{solution}(t)$ of the solved problems from 1850 CE to 2000. Exponential growth models fit rather well the different data sets. The best fits to these three data sets give respectively

$$N_{closed}(t) = \exp[0.02*(t-1790)] + 1.9 \tag{8}$$
$$N_{stated}(t) = \exp[0.02*(t-1875)] + 2.3 \tag{9}$$
$$N_{solution}(t) = \exp[0.035*(t-1902)] - 6.5 \tag{10}$$

where t is given in unit of years and is counted since the beginning of the present era.

The average growth rate of the number of new conjectures is approximately equal to 0.02 year$^{-1}$, corresponding to a tripling of the number of new conjectures every 55 years. This growth rate is close to the average growth rate of the World human population over the same period, as shown in Fig. 3. The best exponential fit to the World population from 1750 to present (data retrieved from the United Nations website [6]) is

$$N(t) = \exp[0.018*(t-1905.6)] + 0.552 \tag{11}$$

Taking the growth of the World population as a proxy for the growth of the number of mathematicians (an assumption which is likely underestimating the true number of mathematicians), we can see that the average growth rate of the number of new conjectures is closely tied to the increase of the population of mathematicians. This average exponential law (11) is only a first order approximation, as it is well known that the growth rate of the World population has varied significantly over the last few centuries [7-9]. However, given the coarse-grained nature of our dataset on mathematical conjectures, the average exponential growth (6) provides a reasonable first representation of the non-stationarity resulting from the increase of the population of mathematicians.

## 5 The consequence of increasing birth flow

The exponential growth of the birth flow of conjectures has one important side effect. It implies that the observable distribution of times-to-proof is bounded by an exponential distribution with rate 0.02. In other words, it cannot decay slower asymptotically than an exponential with rate 0.02. In terms of a CCDF plot, this implies that the empirical times-to-proof distribution of closed and open problems have to lie to the left of this exponential distribution. The reason for this can be seen by writing the distribution $P(\tau)$ of waiting times between formulation and proof in terms of the rate $r(t_1) = r_0 e^{at_1}$ of conjecture formulations and of the conditional distribution $p(t_2|t_1)=f(t_2-t_1)$ that the conjecture will be proved at $t_2$ given that it has been formulated at $t_1$. We assume a constant growth rate $a$ for $r(t_1)$ and stationarity for $p(t_2|t_1)$. This second condition provides an upper bound for the distribution. In other words, the true distribution will decay at least as fast as derived from the assumption of stationarity of $p(t_2|t_1)$. We have

$$P(\tau) = \int_0^t dt_1 \int_0^t dt_2 r(t_1) p(t_2|t_1) \delta(t_2 - t_1 - \tau) = r_0 \int_0^t d_2\, e^{a(t_2-\tau)} f(\tau) = C e^{-a\tau} f(\tau) \quad (12)$$

where

$$C = r_0 \int_0^t dt_2\, e^{at_2} \quad (13)$$

Thus, $P(\tau)$ decays no slower than $e^{-a\tau}$, that is, proportional to the inverse of the rate of conjecture births.

Fig. 4 plots the CCDF of closed and open conjectures defined by equation (1)

and (2), together with the above derived exponential bound. Comparing the three distributions, we can see clearly that the CCDF of the open conjectures lies above the exponential bound. This implies that, although we have used a better and extended dataset, a significant number of conjectures are likely to be still missing in the bulk of the distribution. In other words, there are many missing conjectures with intermediate values of their time-to-proof, compared with the conjectures with extremely large waiting times. Fig. 5 depicts the empirical distribution obtained by removing all times-to-proof smaller than 20 years, i.e., by introducing a lower threshold, so that we can hope that the number of missing conjectures is reduced and the data is less incomplete. We can see that the time-to-proof of closed and open conjectures lie just on the exponential bound, except for the four largest data points (we shall come back later to the status of these outliers). Pushing upward the threshold above 20 years does not change the basic behavior that both CCDF's lie on the exponential bound. Using this theory, the fact that the two distributions coincide confirms that the underlying distribution of time-to-proof is asymptotically close to an exponential distribution.

## 6. Simulation analysis

We now attempt to find distributions of time-to-proof and their associated parameters that could generate the distributions shown in fig. 5. We first generate a set of $N$ instants $t_1^i$, $i=1, ..., N$, corresponding to the formulation times of $N$ mathematical problems. These $N$ times are sampled according to a Poisson process with an intensity growing exponentially with the rate 0.02 year$^{-1}$, as obtained from the fits shown in fig 2. This generation of conjectures mimics the structure of our dataset as described above. This reflects a scenario in which each mathematician generates on average the same number of conjectures per unit time, while the number of mathematicians increases roughly exponentially in parallel with the growth of the human population. Then, a naïve approach would go as follows. For each conjecture inception time $t_1^i$, we draw a random number $\tau_t^c$ corresponding to the time-to-proof of this conjecture. This random number $\tau_t^c$ is generated by using an intrinsic distribution associated with the way mathematics would be practiced by a population of mathematicians of constant size, technology and mental prowess.

We have constructed synthetic catalogues of conjectures with their birth and proof times, using four different families of distributions, namely exponential, lognormal, inverse Gaussian and Burr type-III distribution. For each of these four families of distributions, the parameters were set so as to fit the empirical distribution as closely as possible and, at the same time, to reproduce the ratio of the number of closed to open conjecture (i.e. 42:80). For the exponential family, we find that the rate $\lambda = 0.01$/year provides the best fit. For the lognormal distribution, the best parameters correspond to a log-average of $\mu = 4.2$, and standard deviation $\sigma = 1.2$ For the inverse Gaussian distribution ,

$$f_{ig}(x) = [\frac{\lambda}{2\pi x^3}]^{0.5} \exp\frac{-\lambda(x-\mu)^2}{2\mu^2 x}$$

the best parameters are $\lambda = 170$, $\mu = 72.25$. For the Burr distribution

$$pdf_{BurrIII}(x) = \frac{cd}{x^{c+1}(1+x^{-c})^{d+1}}$$

the best parameters are $c = 0.5$, $d = 3$. As was expected from the result of the previous section, the goodness of fit of the distributions seems to be slightly better for distributions that asymptotically behave like an exponential distribution rather than a power-law distribution. For this reason, we show only the distributions for the exponential family together with the empirical distributions in figure 6.

However, we find that the differences in fit with the catalogues generated with the other distributions are not significant. The paucity of our dataset makes it nearly impossible to distinguish whether distribution "A" provides a better fit than distribution "B". Using the exponential distribution corresponds to following Occam's razor of parsimony with the simplest model providing the best fit. This suggests that most of the conjectures in our database can be described approximately by an underlying waiting time-to-proof distribution that is an exponential distribution with rate 0.01/year.

## 7. Concluding remarks

Motivated by the possibility of measuring an evolution of the production of mathematicians over history and of predicting the waiting time till the proof of a challenging conjecture such as the P-versus-NP problem, we have analyzed the distribution of time-to-proof of the best available general catalogue of mathematical conjectures. As we have dug into the statistical analysis, we have realized the need to take into account the severe non-stationarity of the problem.

As illustrated by figure 6, it is clear that our best model is not the whole story. In particular, half-a-dozen closed conjectures depart rather significantly from the proposed best model. One possible reason for this deviation is that the assumption of an exponential growth of the rate of conjecture births may be too simplified, due to the known deviation of the human population growth from a simple exponential process [6-9]. Using a more realistic birth flow model of conjectures is necessary before we may hope to observe the impact of a possible increase in mathematical productivity.

Moreover, we cannot exclude the existence of incompleteness of the available dataset, in particular of the likely severe under-sampling of the many conjectures whose time-to-proof are in the range from years to a few decades. Only conjectures that have resisted mathematician assaults and/or played particularly distinguished meaningful roles in the structure and/or history of mathematics are likely to acquire the status and the fame to be recorded in databases such as the one we have used. These remarks illustrate the difficulties one is generally confronted with when attempting to extract the distribution of time-to-proof that would really reveal the intrinsic productivity of mathematicians over history.

To conclude, notwithstanding all these difficulties and caveats, if we have to make a best guess and revisit the question first raised by Arbesman and Courtland, 2010), we can use the exponential distribution with rate 0.01/year together with the exponential growth of the mathematician population to calculate the probability for the "P versus NP problem" to be solved by the year 2024. We obtain the value 41%, which suggests that the original estimation of a 50% chance [1] was rather optimistic but still of the right order of magnitude

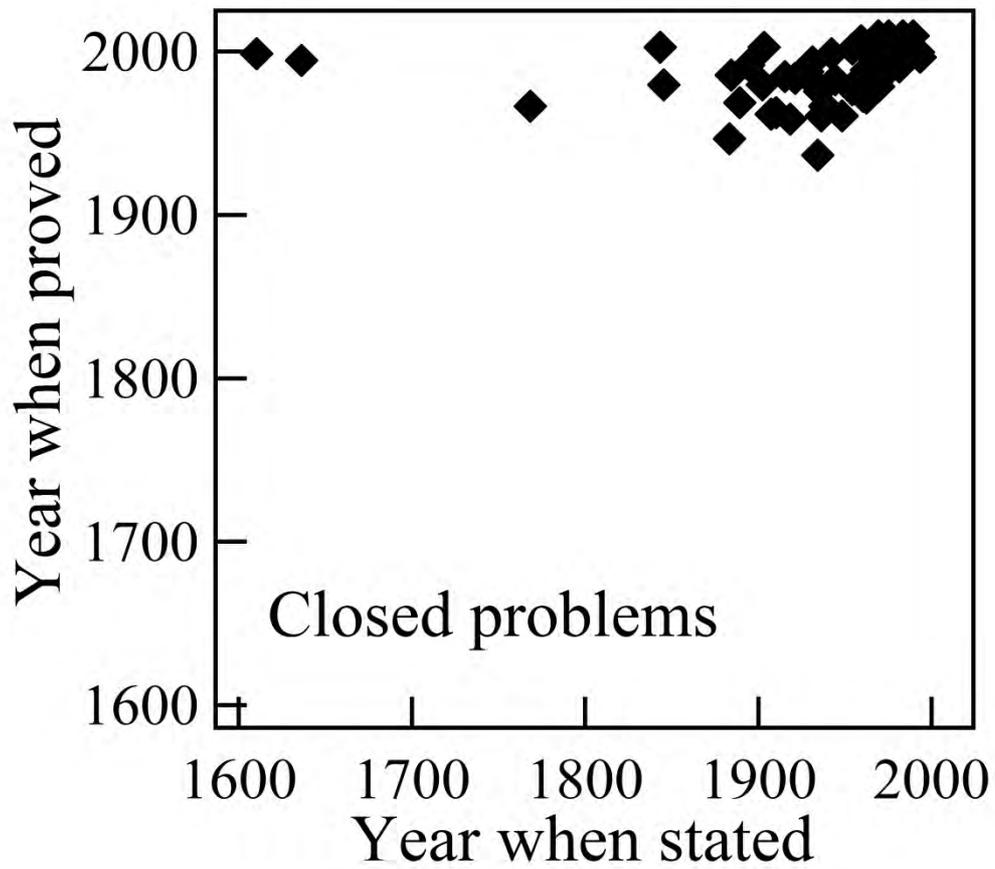

Fig. 1: Scatter plot showing when a mathematical problem was formed and when it was resolved for all 60 closed problems in our dataset, starting from the year 1600.

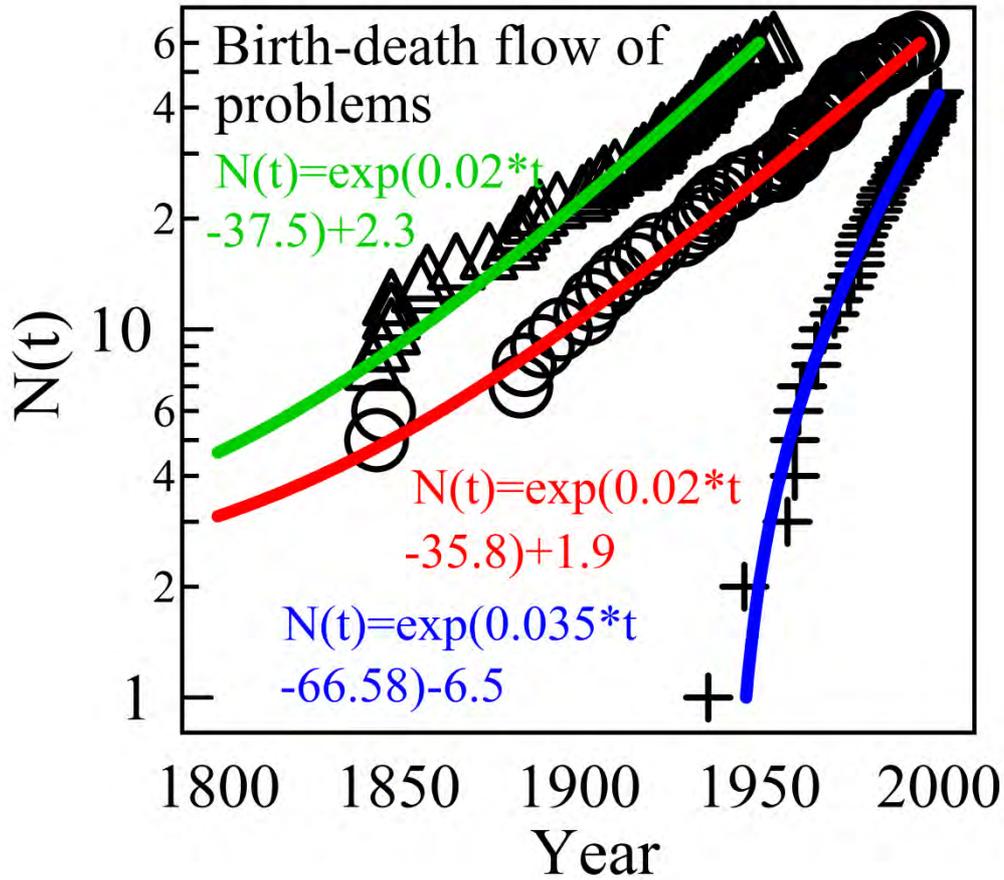

Fig. 2: Cumulative number of stated problems $N_{closed}(t)$ that have found a solution (circles), cumulative number of stated problems $N_{stated}(t)$ that are both closed and still open (triangles) and the cumulative number $N_{solution}(t)$ of the solutions of the solved problems (crosses) from 1850 CE to 2000. The continuous lines correspond to the exponential growth models (8-10).

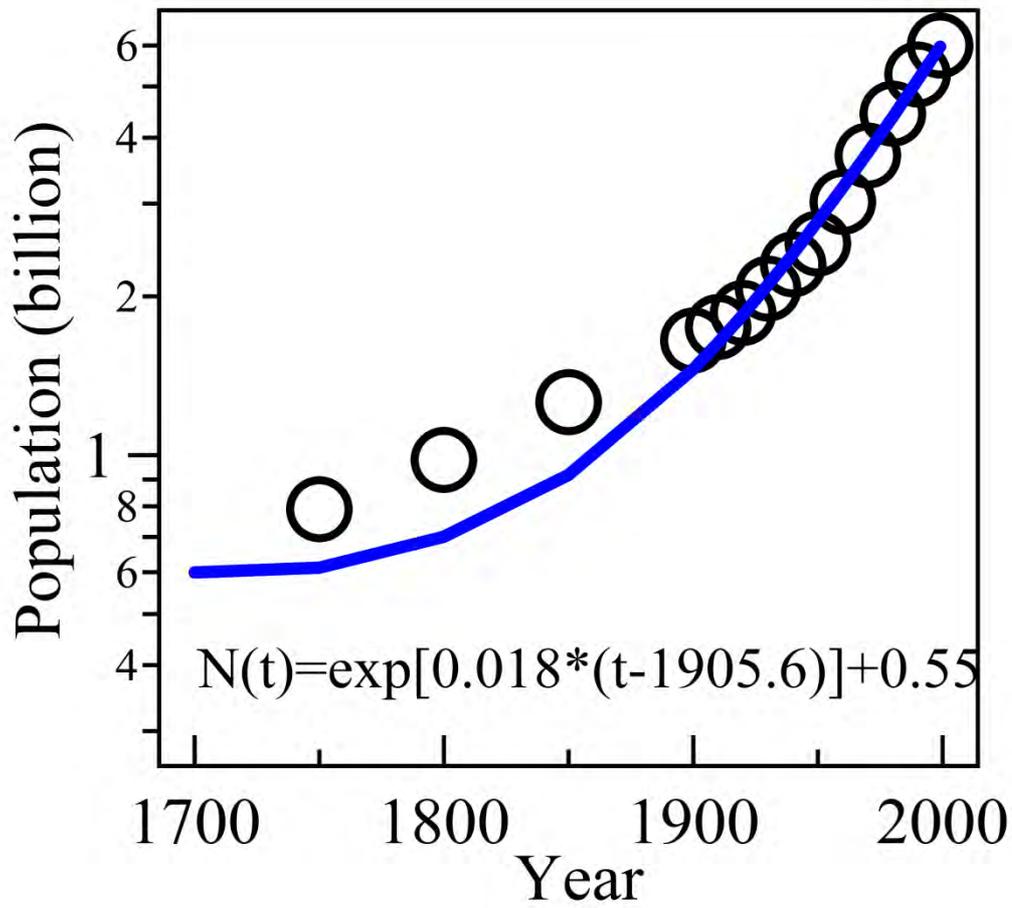

Fig. 3: Growth of the World population from 1750 to present (data retrieved from the United Nations website [6]), taken as the simplest proxy for the growth of the relevant population of mathematicians. This growth can be reasonably approximated by an exponential growth given by expression (11).

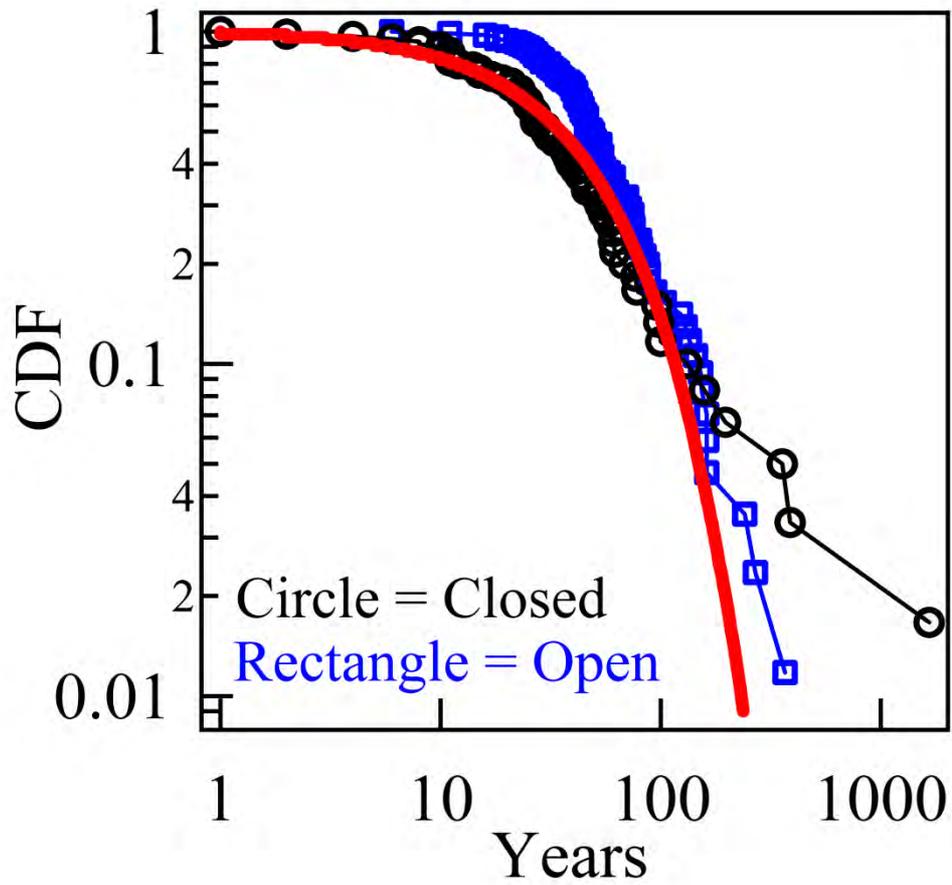

Fig. 4: Complementary cumulative distribution functions (ccdf) of the times-to-proof for open problems (rectangles), closed problems (circles) and an exponential distribution with rate 0.02 (continuous line).

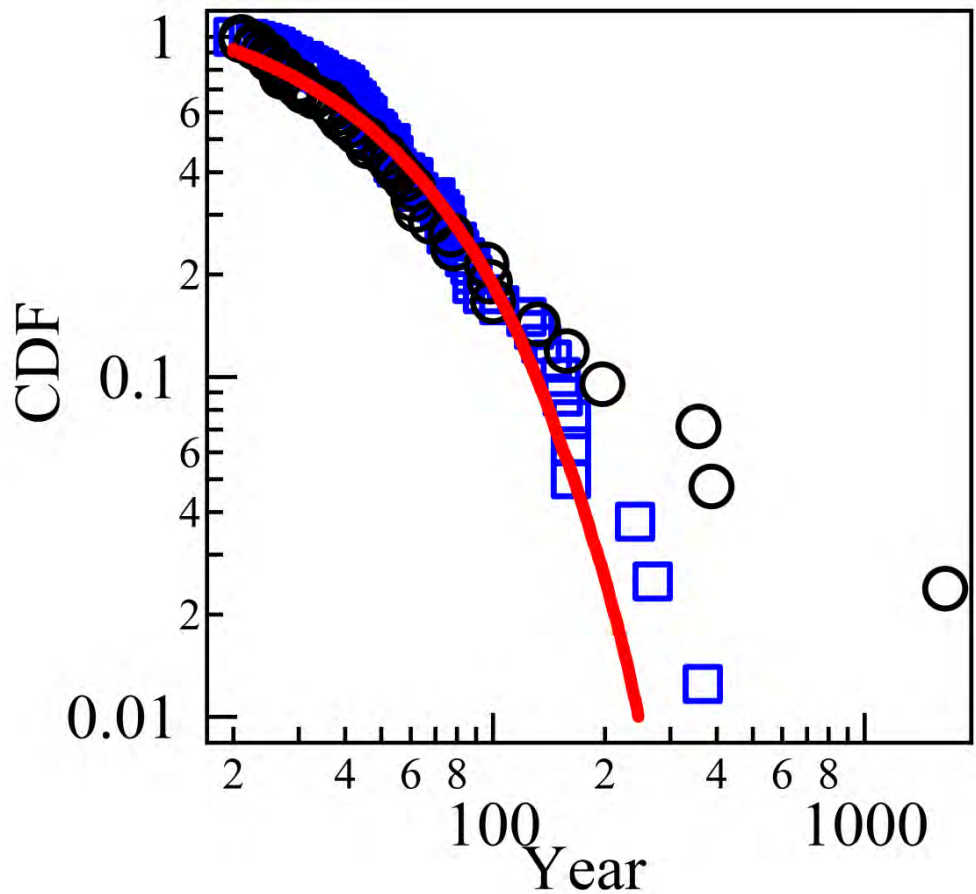

Fig. 5: Complementary cumulative distribution functions (ccdf) of the times-to-proof for open problems (rectangles), closed problems (circles) and an exponential distribution with rate 0.02 (continuous line), obtained by introducing a lower threshold equal to 20 years, i.e., by removing all times-to-proof smaller than 20 years.

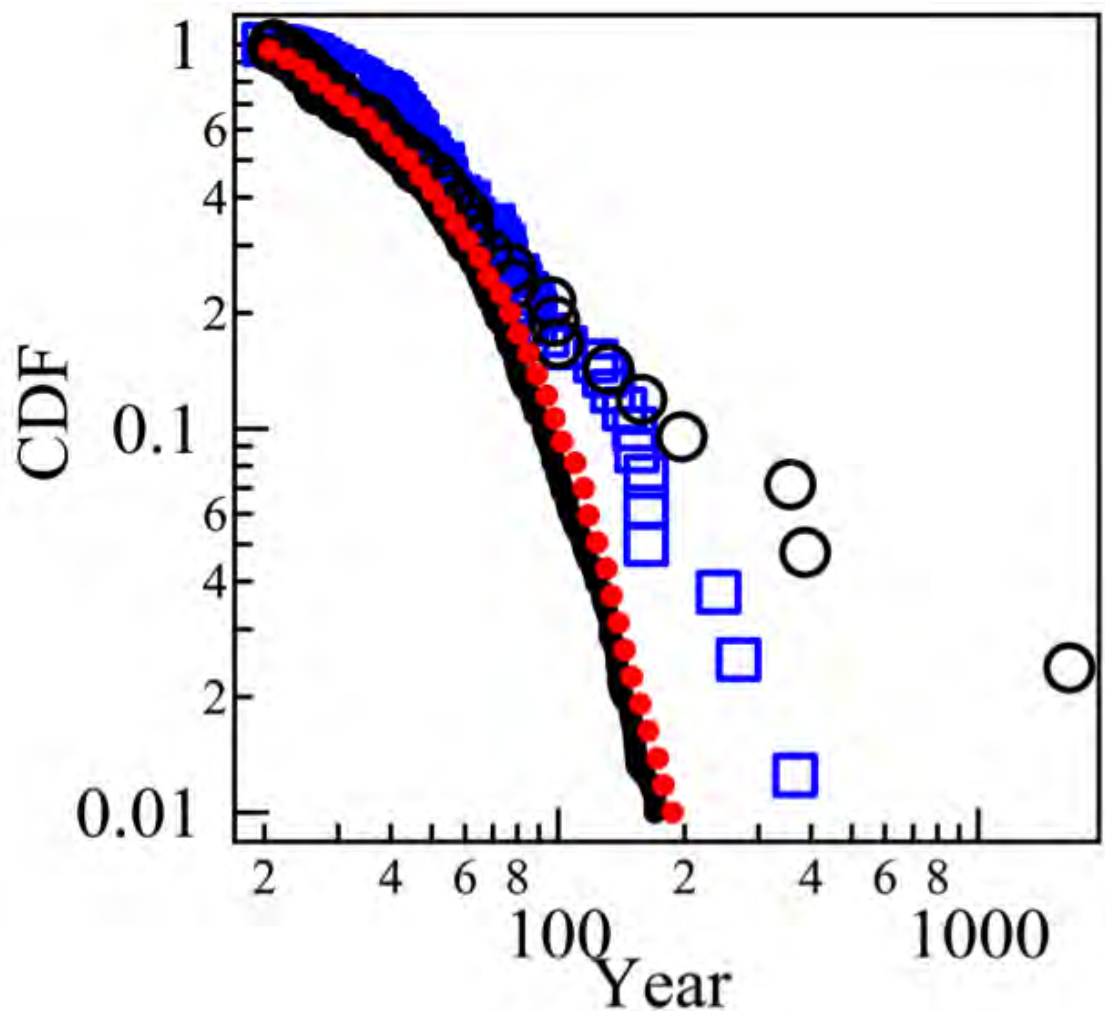

Fig .6: Comparison between the empirical cumulative distribution functions (ccdf) of the times-to-proof for open problems (rectangles), closed problems (circles) and the ccdf generated by the simple model presented in the text using an exponential distribution of the intrinsic time-to-proofs with rate $\lambda = 0.01$/year.

[End]